\documentclass[12pt]{article}
\usepackage{graphicx} 
\usepackage{tensor}
\usepackage{amssymb,amsmath,mathrsfs,amsthm, hyperref,tensor, comment, bm}
\usepackage{mathtools}
\usepackage[style=ext-authoryear-comp,maxcitenames=2,uniquename=false,backend=biber,uniquelist=false,articlein=false]{biblatex}
\DeclareNameAlias{sortname}{family-given}
\usepackage[T1]{fontenc}

\usepackage[usenames,dvipsnames,monochrome]{xcolor}

\usepackage[textsize=tiny]{todonotes}

\DeclareCiteCommand{\citeyear}
    {}
    {\bibhyperref{\printdate}}
    {\multicitedelim}
    {}

\hypersetup{
	colorlinks=true,            
        breaklinks=true,
	linkcolor=Black,
	citecolor=MidnightBlue,
	urlcolor=MidnightBlue            
 }

\newtheorem{prop}{Proposition}
\newtheorem{definition}{Definition}

\title{A primer on Carroll gravity}
\author{Eleanor March\footnote{Faculty of Philosophy, University of Oxford. eleanor.march@philosophy.ox.ac.uk}~ \& James Read\footnote{Faculty of Philosophy, University of Oxford. james.read@philosophy.ox.ac.uk  (Corresponding author.)}}
\date{}

\begin{document}

\maketitle

\begin{abstract}
    The ultra-relativistic limit of general relativity is Carroll gravity. In this article, we provide (i) a rigorous and thorough exposition of the geometric formalism of {\color{blue}the `magnetic' version of} Carroll gravity, (ii) a presentation of {\color{blue}this} theory as a limit of general relativity in a geometrical, `lightcone-narrowing' sense, and (iii) an exploration of some of the various conceptually interesting features of {\color{blue}this version of} Carroll gravity. 
\end{abstract}


\section{Introduction}\label{sec:intro}

It's well known that Galilei group $\text{Gal} (d) := \left( \text{SO}(d) \ltimes \mathbb{R}^d \right) \ltimes \mathbb{R}^{d+1}$, obtained in the non-relativistic limit of the Poincaré group (heuristically, as $c \rightarrow \infty$), acts on spacetime coordinates $x^\mu = (t, \vec{x} )$ as
\begin{align}
\begin{aligned}
    t' &= t + b, \\
    \vec{x}' &= R \vec{x} + \vec{v}t + \vec{a}.
\end{aligned}
\end{align}
If one instead takes the \emph{ultra}-relativistic limit of the Poincaré group (heuristically, $c \rightarrow 0$) one arrives at the \emph{Carroll group} $\text{Car} (d) := \left( \text{SO}(d) \ltimes \mathbb{R}^d \right) \ltimes \mathbb{R}^{d+1}$. Although this group is the same as that for the Galilei group (see \textcite[ch.\ 2]{HansenThesis} for a detailed discussion), thinking about obtaining this group of transformations via the ultra-relativistic limit means that it acts differently on spacetime coordinates $x^\mu = (t, \vec{x} )$---in particular, it acts as
\begin{align}\label{eq:carrolltransfo}
\begin{aligned}
    t' &= t + \vec{v} . R \vec{x} + b, \\
    \vec{x}' &= R \vec{x} + \vec{a}.
\end{aligned}
\end{align}
This yields different physics, as we'll see. Geometrically, just as Galilean spacetime (the isometry group of which is the Galilei group) can be obtained by `widening' the lightcones of Minkowski spacetime, so too can Carroll spacetime (the isometry group of which is the Carroll group) be obtained by `narrowing' the lightcones of Minkowski spacetime. Carroll gravity in its full glory stands to such a flat-spacetime ultra-relativistic theory as general relativity stands to special relativity (the latter, of course, being set in Minkowski spacetime). {\color{teal} (In this article, when we speak of a `non-relativistic spacetime', we always mean a classical spacetime (e.g.\ Galilean spacetime), in the sense of \textcite[ch.\ 4]{Malament2012-MALTIT}. Of course, there is a more general sense in which Carroll spacetimes are also non-relativistic, but we won't use the word in that more liberal sense.)}


The nomenclature `Carroll gravity' is due to \textcite{Lévy1965}, who wrote this:
\begin{quote}
    In reality, the paradoxical aspects of `Carrollian' invariance come from the fundamental condition [$\Delta x / \Delta t \gg 1$] for the validity of this approximation of relativistic invariance. The laws of the Carrollian transformation [\eqref{eq:carrolltransfo}] can in fact by hypothesis only be applied to large space-like intervals. But two events separated by such an interval are obviously totally causally disconnected [...] We can thus predict that at the corresponding non-relativistic limit the notion of causality will lose almost all content. Indeed, as we see from the formulae [\eqref{eq:carrolltransfo}], by changing the appropriate reference system, we can modify at will the time interval between two events and
in particular change its sign, except in the case where the spatial interval between the two events is zero. In other words, the causal shadow of a given event is reduced to the very place where this event takes place, for any time. This is also perfectly visible [...] when we see in the Carrollian limit the cones `absolute future' and `absolute past' contracting on the axis of time, the region of `absolute elsewhere' invading all space-time. Finally, note that in Galilean theory the time interval between two events is an invariant; conversely, here it is the length of the spatial interval which is invariant. \parencite[p.\ 11, our translation]{Lévy1965}
\end{quote}
One of the central points made by Lévy-Leblond in this passage is that there is a very attenuated notion of causality in Carroll spacetime, because almost \emph{any} two events are elsewhere-related in such a  setting. Another is that the time interval between two events is not an invariant in Carroll spacetime, unlike in the more familiar Galilean spacetime.
The name `Carroll gravity' is {\color{blue}a reference} to Lewis Carroll's \emph{Through the Looking-Glass}, the world of which often seems to lack well-defined time intervals and sensible notions of causality.\footnote{ E.g.: \begin{quote} ``Well, in our country,'' said Alice, still panting a little, ``you’d generally get to somewhere else if you run very fast for a long time, as we’ve been doing.''

``A slow sort of country!'' said the Queen. ``Now, here, you see, it takes all the running you can do, to keep in the same place. If you want to get somewhere else, you must run at least twice as fast as that!'' \end{quote}

Or as Jay-Z later put it:

\begin{quote}
That’s called the Red Queen’s Race \\
You run this hard just to stay in place \\
Keep up the pace, baby \\
Keep up the pace.
\end{quote} }

Before we proceed any further, we should contextualise our work. The first point to make here is this: there are already a few good entry-points for the study of Carroll gravity in the literature---see, in particular, \textcite{Hartong}, \textcite[Appendix]{Bekaert:2015xua}, and \textcite{Niedermaier}. However, those articles either stress the gauge-theoretic origins of Carroll gravity, or proceed at a high level of mathematical abstraction. {\color{teal} (A gauge-theoretic version of magnetic Carroll gravity is given by \textcite{Bergshoeff}; we discuss below the difference between magnetic and electric Carroll gravity. Other important recent works on Carroll gravity include (i) \textcite{pekar, FOF}, which investigate Carroll gravity from the point of view of Cartan geometry, (ii) \textcite{Guerrieri}, which investigates the ultra-relativistic limit of Lovelock--Cartan gravity, and (iii) \textcite{Ciambelli}, which explores the structure of Carroll geometries from the fibre bundle point of view.)} Of course, such approaches are of great value---however, our goal in this article is to complement that work, by offering a presentation of Carroll gravity roughly at the level of \textcite{Wald:1984rg}, or \textcite[ch.\ 4]{Malament2012-MALTIT}. In doing this, we will prove a number of propositions about the theory and its status as the ultra-relativistic limit of general relativity which (to our knowledge) have not appeared in the literature up to this point. As such, we intend the article to be a valuable point of entry into Carroll gravity. The second point to make by way of contextualising our article is this: it's by now relatively well-known in literature on Carroll gravity that there in fact exist two different limits of the theory---an `electric' limit and a `magnetic' limit---depending upon how one scales the relevant objects (such as the metric) when taking the ultra-relativistic limit of general relativity. (The terminology of `electric' and `magnetic' versions of Carroll gravity was introduced by \textcite{Henneaux1}.) The `electric' ultra-relativistic limit of general relativity yields a theory which in general has torsion, and which is closely associated with `Type II' limits (which have become better-known recently in the context of taking the \emph{non}-relativistic limit of general relativity)---see e.g.\ \textcite{HansenThesis} for further background. In this article, we'll concern ourselves only with the \emph{magnetic} ultra-relativistic limit, and will defer discussion of \emph{inter alia} torsionful ultra-relativistic spacetimes for future work. It bears stressing, then, that what this article offers is a primer on \emph{magnetic} Carroll gravity; this allows us to present results in a pedagogical and (we hope) elegant way, albeit of course at the cost of full generality.

The plan for the article, then, is this. We'll return later, in \S\ref{sec:discussion}, to a conceptual appraisal of Carroll gravity. Before doing so, however, in  \S\ref{sec:Carrollspacetime} we first provide a rigorous, coordinate-independent presentation of the theory (to repeat, in the style of e.g.\  \textcite[ch.\ 4]{Malament2012-MALTIT}), and then discuss in detail in \S\ref{sec:limit} how Carroll gravity can be understood as the ultra-relativistic limit of a relativistic spacetime theory in a geometrical, `lightcone-narrowing' sense.

\section{Carroll spacetimes and Carroll gravity}\label{sec:Carrollspacetime}

Let $M$ be a differentiable four-manifold (assumed connected, Hausdorff, and paracompact). A \textit{Carroll spacetime} is a structure $\langle M,  h_{ab}, s^{ab}, \nabla\rangle$, where $h_{ab}$ is a smooth, symmetric field with signature $(0,1,1,1)$, $s^{ab}$ is a smooth, symmetric field with signature $(1,0,0,0)$ satisfying $s^{ab} h_{bc} = \bm{0}$, and $\nabla$ is a derivative operator satisfying $\nabla_a s^{bc}=\bm{0}$ and $\nabla_ah_{bc}=\bm{0}$. (Throughout this article, we will assume that $\nabla$ is torsion-free, which---to repeat---is to place us in the `magnetic' sector of the theory.) As in the non-relativistic case, we will refer to the condition $s^{ab} h_{bc} = \bm{0}$ as the orthogonality condition and the conditions $\nabla_a s^{bc}=\bm{0}$ and $\nabla_ah_{bc}=\bm{0}$ as the compatibility conditions.

The signature condition for $h_{ab}$ is the requirement that, for every $p \in M$, the tangent space there have a basis $\overset{1}{\zeta}{}^a , \ldots , \overset{4}{\zeta}{}^a$ such that, for all $i, j \in \left\{ 1, 2, 3, 4 \right\}$, $h_{ab} \overset{i}{\zeta}{}^a \overset{j}{\zeta}{}^b = 0$ if $ i \neq j$, and
\begin{equation}
   h_{ab} \overset{i}{\zeta}{}^a \overset{j}{\zeta}{}^b = \begin{cases} \text{$0$ if $i=1$} \\ \text{$1$ if $i = 2,3,4$.} \end{cases}
\end{equation}
{\color{teal}(This carries over straightforwardly from the notion of a signature in a non-relativistic spacetime context, as presented by \textcite[pp.\ 249--50]{Malament2012-MALTIT}.)} Likewise, the signature condition for $s^{ab}$ is the requirement that, for every $p \in M$, the cotangent space there have a basis $\overset{1}{\sigma}{}_a , \ldots , \overset{4}{\sigma}{}_a$ such that, for all $i, j \in \left\{ 1, 2, 3, 4 \right\}$, $s^{ab} \overset{i}{\sigma}{}_a \overset{j}{\sigma}{}_b = 0$ if $ i \neq j$, and
\begin{equation}
   s^{ab} \overset{i}{\sigma}{}_a \overset{j}{\sigma}{}_b = \begin{cases} \text{$1$ if $i=1$} \\ \text{$0$ if $i = 2,3,4$.} \end{cases}
\end{equation}

At any point, we can find a vector $\xi^a$ such that $s^{ab} = \xi^a \xi^b$. We say that a Carroll spacetime $\langle M,  h_{ab}, s^{ab}, \nabla\rangle$ is temporally orientable just in case there exists a continuous (globally defined) vector field $\xi^a$ that satisfies this decomposition condition at every point. Going forward, we will generally assume temporal orientability and so consider Carroll spacetimes of the form $\langle M,  h_{ab}, \xi^a, \nabla\rangle$. In such spacetimes, we have the orthogonality condition $\xi^a h_{ab} = \bm{0}$ and the compatibility condition $\nabla_a \xi^b = \bm{0}$. From the compatibility conditions there follows the condition $\pounds_\xi h_{ab}=\bm{0}$, which is the analogue of the $\mathrm{d}_at_{b}=\bm{0}$ condition in non-relativistic spacetimes (see \textcite[p.\ 251]{Malament2012-MALTIT}) and which renders meaningful absolute spatial relations between all events in Carroll spacetimes; {\color{teal} we return to this in \S\ref{sec:discussion}. (Here and throughout, $\pounds$ denotes the Lie derivative.)}

The objects $\xi^a$ and $h_{ab}$ induce a classification of vector fields as timelike or spacelike. We will say that a vector field $v^a$ is timelike iff $h_{an}v^n=\bm{0}$ and spacelike otherwise (from this it follows that $\xi^a$ itself is timelike). Similarly, $\xi^a$ and $h_{ab}$ also induce a classification of covector fields as spacelike or timelike: a covector field $\sigma_a$ is spacelike if $\sigma_n\xi^n=0$ and timelike otherwise. We will say that a covector $t_a$ is unit timelike iff $t_n\xi^n=1$.

We also have the following (equivalent) characterisation of timelike and spacelike vector and covector fields:
\begin{prop}\label{prop:spaceliketimelike}
    Let $\langle M, h_{ab}, \xi^a, \nabla\rangle$ be a Carroll spacetime. Then the following statements hold:
    \begin{itemize}
        \item[(i)] For any vector field $v^a$, $h_{an}v^n=\bm{0}$ iff $v^a=\alpha\xi^a$ for some $\alpha$.
        \item[(ii)] For any covector field $\sigma_a$, $\sigma_n\xi^n=0$ iff $\sigma_a=h_{an}\sigma^n$ for some $\sigma^n$. 
    \end{itemize}
\end{prop}
\begin{proof}
    The `if' directions of the proposition are immediate. For the `only if' directions, consider first (i). Let $v^a$ be a vector field, let $p\in M$, and let $\overset{i}{\zeta}\vphantom{\zeta}^a$, $i=1,2,3,4$ be a basis for $h_{ab}$ at $p$ in the sense discussed above (so, in particular), $\overset{1}{\zeta}\vphantom{\zeta}^a=\xi^a(p)$. Then $v^a(p)=\alpha\xi^a(p)+\overset{2}{\alpha}\overset{2}{\zeta}\vphantom{\zeta}^a+\overset{3}{\alpha}\overset{3}{\zeta}\vphantom{\zeta}^a+\overset{4}{\alpha}\overset{4}{\zeta}\vphantom{\zeta}^a$ for some $\alpha$, $\overset{i}{\alpha}$. Thus if $h_{an}v^n=\bm{0}$, it follows that $v^a(p)=\alpha(p)\xi^a(p)$ at each $p\in M$ and hence that $v^a=\alpha\xi^a$. The `only if' direction of (ii) follows by dimensionality considerations. 
\end{proof}
Note that unlike in the case of non-relativistic spacetimes, the spacelike vectors at some $p\in M$ do not form a vector space. The spacelike covectors at a point, however, do.

Analogously with the case of a non-relativistic spacetime, the inverse of the degenerate metric $h_{ab}$ is non-unique; however, one can specify an inverse relative to an arbitrary unit timelike covector field:

\begin{prop}
    Let $\langle M, h_{ab}, \xi^a, \nabla\rangle$ be a Carroll spacetime and let $t_a$ be an arbitrary unit timelike covector field. Then there is a (unique) smooth, symmetric field $\hat{h}^{ab}$ on $M$ satisfying the conditions:
    \begin{align}
        t_a \hat{h}^{ab} &= \bm{0}, \\
        h_{ab} \hat{h}^{bc} &= \delta\indices{^c_a} - t_a \xi^c =: \hat{h}\indices{^{c}_{a}}.
    \end{align}
\end{prop}
\begin{proof}
    We can define a symmetric field $\hat{h}^{ab}$ by specifying its action on $t_a$ and on an arbitrary spacelike covector field $\mu_a$. So consider the field $\hat{h}^{ab}$ which annihilates the former (thereby satisfying the first of the above conditions) and which makes the assignment
    \[ \hat{h}^{ab} \mu_b = \sigma^a - \xi^a ( t_c \sigma^c),\] where $\sigma^a$ is any vector such that $\sigma^a h_{ab} = \mu_b$; the particular choice of $\sigma^a$ plays no role here. From the way that we have defined $\hat{h}^{ab}$, the second of the above conditions then follows.
\end{proof}

{\color{blue} (The globally-defined timelike $\xi^a$, $t_a$ with $\xi^a t_a = 1$, define a `Carroll structure' and $t_a$ is the associated Ehresmann connection. However, the setting which we consider in this paper is more general, as we take it that $t_a$ is not part of the spacetime structure, i.e.\ we don't have any particular privileged $t_a$ or collection thereof. Our thanks to an anonymous referee for inviting us to say something on Carroll structures here.)} In analogy with \textcite[p.\ 255]{Malament2012-MALTIT}, we can call $\hat{h}^{ab}$ the `spatial projection field' relative to $t_a$.

It is also worth dwelling further on the significance of the temporal compatibility condition. We have the following proposition:
\begin{prop}
    Let $\langle M, h_{ab}, \xi^a, \nabla\rangle$ be a Carroll spacetime. Then parallel transport of timelike vectors is (at least locally) path-independent.
\end{prop}
\begin{proof}
    Let $p\in M$ and $\overset{p}{v}\vphantom{v}^a$ be an arbitrary timelike vector at $p$. To show that parallel transport of timelike vector fields on $M$ is (at least locally) path independent, it will suffice to show that for some open set $O$ containing $p$, there is an extension of $\overset{p}{v}\vphantom{v}^a$ to a vector field $v^a$ on $O$ which is constant, i.e.~$\nabla_av^b=\bm{0}$. By proposition \ref{prop:spaceliketimelike}, it follows that $\overset{p}{v}\vphantom{v}^a=\overset{p}{\alpha}\xi^a(p)$ for some $\overset{p}{\alpha}$. Since $\nabla_a\xi^b=\bm{0}$, we can find a constant vector field extending $\overset{p}{v}\vphantom{v}^a$ to some open region $O$ containing $p$ if there exists a scalar field $\alpha$ on $O$ satisfying $\nabla_a\alpha=\bm{0}$ and which agrees with $\overset{p}{\alpha}$ at $p$ (we just take $v^a=\alpha\xi^a$). But since $\nabla$ is torsion-free, constant scalar fields always exist locally. 
\end{proof}

Let's consider next the extent to which the derivative operator in a Carroll spacetime is fixed uniquely by its compatibility with these objects. {\color{teal} Before doing so, however, we should recall some notation. Throughout this article, $C\indices{^{a}_{bc}}$ denotes the `difference tensor' between two derivative operators $\nabla$ and $\nabla'$, which is defined as the smooth symmetric tensor field on $M$ satisfying the following condition for all smooth tensor fields $\alpha^{a_1 \ldots a_r}_{b_1 \ldots b_s}$ on $M$:
\begin{multline}\label{differencetensor}
    \left( \nabla'_m - \nabla_m \right) \alpha^{a_1 \ldots a_r}_{b_1 \ldots b_s} = \alpha^{a_1 \ldots a_r}_{n b_2 \ldots b_s} C\indices{^{n}_{m b_1}} + \cdots + \alpha^{a_1 \ldots a_r}_{b_1 \ldots b_{s-1} n} C\indices{^{n}_{m b_s}} \\ - \alpha^{n a_2 \ldots a_r}_{b_1 \ldots b_s} C\indices{^{a_1}_{mn}} - \cdots - \alpha^{a_1 \ldots a_{r-1} n}_{b_1 \ldots b_s} C\indices{^{a_r}_{mn}}.
\end{multline}
(See \textcite[Proposition 1.7.3]{Malament2012-MALTIT}.) For $\nabla$, $\nabla'$ and $C\indices{^{a}_{bc}}$ satisfying \eqref{differencetensor}, we follow \textcite[p.\ 53]{Malament2012-MALTIT} in adopting the shorthand $\nabla' = \left( \nabla , C\indices{^{a}_{bc}} \right)$.}

{\color{blue} With this in mind, the extent to which uniqueness of the derivative operator in a Carroll spacetime} breaks down is characterised by the following proposition:

\begin{prop}
    Let $\langle M, h_{ab}, \xi^a, \nabla\rangle$ be a Carroll spacetime and let $\Tilde{\nabla}$ be any second derivative operator on $M$. Then $\Tilde{\nabla}$ is compatible with $\xi^a$ and $h_{ab}$ iff $\Tilde{\nabla}=(\nabla, \xi^a\kappa^{nm}h_{bn}h_{cm})$ for some symmetric $\kappa^{ab}$.
\end{prop}
\begin{proof}
    For the `if' direction, assume that $\Tilde{\nabla}$ has the stated form. Since $\nabla$ is compatible with $\xi^a$ and $h_{ab}$, it follows from orthogonality of the metrics that 
    \begin{equation*}
        \Tilde{\nabla}_a\xi^b=\nabla_a\xi^b-C\indices{^b_a_n}\xi^n=\xi^b\kappa^{rm}h_{ar}h_{nm}\xi^n= \bm{0},
    \end{equation*}
    and
    \begin{align*}
\Tilde{\nabla}_ah_{bc}&=\nabla_ah_{bc}+C\indices{^n_a_b}h_{nc}+C\indices{^n_a_c}h_{bn} \\ &=\xi^n\kappa^{rm}h_{ar}h_{bm}h_{nc}+\xi^n\kappa^{rm}h_{ar}h_{cm}h_{bn} \\ &= \bm{0} .
    \end{align*}
    
    For the `only if' direction, assume that $\Tilde{\nabla}$ is compatible with $\xi^a$ and $h_{ab}$. We know that $\Tilde{\nabla}=(\nabla, C\indices{^a_b_c})$. Hence
    \begin{equation}\label{eq:temporalcompatibility}
        \Tilde{\nabla}_a\xi^b=\nabla_a\xi^b-C\indices{^b_a_n}\xi^n=C\indices{^b_a_n}\xi^n=\bm{0},
    \end{equation}
    \begin{equation}\label{eq:spatialcompatibility}
        \Tilde{\nabla}_ah_{bc}=\nabla_ah_{bc}+C\indices{^n_a_b}h_{nc}+C\indices{^n_a_c}h_{bn}=C\indices{^n_a_b}h_{nc}+C\indices{^n_a_c}h_{bn}=\bm{0}.
    \end{equation}
    Now consider $C_{abc}=h_{an}C\indices{^n_b_c}$. It is spacelike, i.e.~contraction on any index with $\xi^a$ yields $\bm{0}$. Moreover, it satisfies the following two conditions {\color{teal} (we obtain these specific results because we follow the index placement conventions of \textcite{Malament2012-MALTIT}---also for Riemann tensor indices, as discussed below)}:
    \begin{align*}
        C_{abc}&=-C_{cba} , \\
        C_{abc}&=C_{acb} ,
    \end{align*}
    where the first comes from \eqref{eq:spatialcompatibility} and the second from the symmetry of $C\indices{^a_b_c}$. Hence
    \begin{equation*}
        C_{abc}=-C_{cba}=-C_{cab}=C_{bac}=C_{bca}=-C_{acb}=-C_{abc},
    \end{equation*}
    so $C_{abc}=\bm{0}$. Now let $t_a$ be a unit timelike covector field (so $t_n\xi^n=1$) and $\hat{h}^{ab}$ the corresponding spatial projector. Then $\bm{0}=\hat{h}^{an}C_{nbc}=(\delta^a_n-t_n\xi^a)C\indices{^n_b_c}$ and hence
    \begin{equation}\label{eq:Ccondition}
        C\indices{^a_b_c}=t_nC\indices{^n_b_c}\xi^a.
    \end{equation}
    Finally, consider $\kappa^{ab}=t_nC\indices{^n_m_r}\hat{h}^{am}\hat{h}^{br}$. It is symmetric, and we claim, satisfies the conditions of the proposition. For this, we compute:
    \begin{align*}
    \xi^{a}\kappa^{nm}h_{bn}h_{cm}&=\xi^{a}t_rC\indices{^r_s_t}\hat{h}^{ns}\hat{h}^{mt}h_{bn}h_{cm}\\
        &=\xi^{a}t_rC\indices{^r_s_t}(\delta^s_b-t_b\xi^s)(\delta^t_c-t_c\xi^t)\\
        &=\xi^{a}t_rC\indices{^r_b_c}\\
        &=C\indices{^a_b_c} ,
    \end{align*}
    where we have made use of \eqref{eq:temporalcompatibility} and the symmetry of $C\indices{^a_b_c}$ in the third equality and \eqref{eq:Ccondition} in the fourth.
\end{proof}
We now consider the curvature tensor $R\indices{^a_b_c_d}$ associated with $\nabla$ and its symmetry properties. {\color{teal} (Recall that $R\indices{^a_b_c_d}$ is defined such that, for all smooth vector fields $v^b$ on $M$, $R\indices{^a_b_c_d} v^b := - 2 \nabla_{[c} \nabla_{d]} v^a$. To repeat: here and throughout this article, we follow the index placement conventions of \textcite{Malament2012-MALTIT}.)} Of course, it satisfies the conditions {\color{teal} (see \textcite[prop.\ 1.8.2]{Malament2012-MALTIT})}:
\begin{align}
    R\indices{^a_{[bcd]}} &=\bm{0}, \label{eq:bianchi1} \\
    R\indices{^a_b_{(cd)}} &=\bm{0},  \label{eq:riemannantisymmetric}
\end{align}
as well as Bianchi's identity,
\begin{equation}
    \nabla_{[m} R\indices{^{a}_{|b|cd]}} = \bm{0}.
\end{equation}
The compatibility conditions $\nabla_a\xi^b=\bm{0}$ and $\nabla_ah_{bc}=\bm{0}$ further imply that 
\begin{align}
    \xi^nR\indices{^a_n_c_d}&=\bm{0}, \label{eq:timelikeriemann} \\
    R\indices{_{(ab)}_c_d}&=\bm{0},\label{eq:spatialantisymmetry}
\end{align}
from which it follows that 
\begin{align}
    R_{abcd}-R_{cdab}&=\bm{0},\\
    \xi^nR\indices{^a_{[bc]n}}&=\bm{0},\label{eq:riemannxiantisymmetric}\\
    \xi^nR\indices{_a_{bcn}}&=\bm{0}.\label{eq:downstairsnewtonian}
\end{align}

Now consider the Ricci tensor $R_{ab}=R\indices{^n_a_b_n}$ and the scalar curvature field $R=\xi^n\xi^mR_{nm}$. Obviously the latter vanishes, and we claim the former is symmetric. To verify this, let $t_a$ be an arbitrary unit timelike covector field. We have
\begin{equation} \label{eq:spatialcontracted}
    R\indices{^n_n_a_b}=\hat{h}^{nm}R\indices{_n_m_a_b}=\bm{0}.
\end{equation}
(Since $\hat{h}^{nm}R\indices{_n_m_a_b}=(\delta\indices{^s_r}-t_r\xi^s)R\indices{^r_s_a_b}=R\indices{^s_s_a_b}$, which together with \eqref{eq:spatialantisymmetry} gives us \eqref{eq:spatialcontracted}.) Thus from \eqref{eq:bianchi1} and \eqref{eq:riemannantisymmetric}
\begin{equation*}
    R_{ab}-R_{ba}=R\indices{^n_a_b_n}-R\indices{^n_b_a_n}=R\indices{^n_a_b_n}+R\indices{^n_b_n_a}=-R\indices{^n_n_a_b},
\end{equation*}
so by \eqref{eq:spatialcontracted} we have $R_{ab}=R_{ba}$. 

We now consider some further properties of the curvature tensor $R\indices{^a_b_c_d}$ associated with $\nabla$ in a Carroll spacetime. Let $\langle M, h_{ab}, \xi^a, \nabla\rangle$ be a Carroll spacetime. Of course, we say that it is flat iff $R\indices{^a_b_c_d}=\bm{0}$. In parallel, we will say that it is \textit{spatially flat} iff $R_{abcd}=\bm{0}$. To motivate this definition, we need to say something about induced derivative operators on spacelike hypersurfaces.

Let $t_a$ be a unit timelike covector field. We will say that a hypersurface $S$ is spacelike relative to $t_a$ iff, for all $p\in S$ and all tangent vectors $\sigma^a$ to $S$ at $p$, $t_n\sigma^n=0$. (Clearly, if $S$ is spacelike relative to $t_a$ then it is also spacelike, i.e.~all smooth curves with images in $S$ are spacelike.) Similarly, we will say that a tensor field is spacelike relative to $t_a$ iff contraction on any of its indices with $t_a$ or $\xi^a$ yields $\bm{0}$. In what follows, let $S$ be a hypersurface which is spacelike relative to $t_a$. {\color{teal} (Of course, such a hypersurface exists (at least locally) iff the distribution of spacelike vectors relative to $t_a$ at each $p\in M$ is integrable, i.e.~$t_{[a}\mathrm{d}_bt_{c]}=\bm{0}$. In what follows, we will assume that this is the case.)} We can think of tensor fields which are spacelike relative to $t_a$ as living on $S$. Clearly $h_{ab}$ and $\hat{h}^{ab}$ both qualify as spacelike relative to $t_a$, as does $\hat{h}\indices{^a_b}$. Note that $h_{ab}$ does not annihilate any non-zero vectors which are spacelike relative to $t_a$ and $\hat{h}\indices{^a_b}$ preserves all vectors which are spacelike relative to $t_a$. In other words, we can think of $h_{ab}$ as a non-degenerate metric on $S$, which induces a unique derivative operator $D$ on $S$. The action of $D$ on a tensor field is given by first acting with $\nabla$ and then projecting all contravariant indices as well as $\nabla$ with $\hat{h}\indices{^a_b}$. So, for example, the action of $D$ on $\alpha\indices{^a_b_c}$ is given by 
\begin{equation}
    D_a\alpha\indices{^b_c_d}=\hat{h}\indices{^b_n}\hat{h}\indices{^m_a}\nabla_m\alpha\indices{^n_c_d}.
\end{equation}
Tensor fields with other index structures are handled analogously. The projection ensures that the resultant field is spacelike relative to $t_a$. There is
no need to project the other covariant indices, since $\nabla_a\xi^b=\bm{0}$. (One can check that $D$ satisfies the conditions to be a derivative operator on $S$, and that $D_ah_{bc}=\bm{0}$ and $D_a\hat{h}^{ab}=\bm{0}$.) 

The following proposition then serves to motivate our above definition of spatial flatness:
\begin{prop}[Spatial flatness]
    Let $\langle M, h_{ab}, \xi^a, \nabla\rangle$ be a Carroll spacetime. Then given any unit timelike covector field $t_a$ and any spacelike hypersurface $S$ relative to $t_a$, $R_{abcd}=\bm{0}$ throughout $S$ iff parallel transport of spacelike covectors within $S$ is (at least locally) path-independent.
\end{prop}
\begin{proof}
    Let $t_a$ be any unit timelike covector field, $S$ a spacelike hypersurface relative to $t_a$, let $\sigma_a$ be any spacelike covector field on $S$, and $\mu^a$ be any vector field which is spacelike relative to $t_a$. Then $\sigma_a$ automatically qualifies as spacelike relative to $t_a$, and $\mu^nD_n\sigma_a=\mu^n\hat{h}\indices{^m_n}\nabla_m\sigma_a=\mu^n\nabla_n\sigma_a$, i.e.~$\nabla$ and $D$ induce the same conditions for parallel transport of spacelike covectors throughout $S$. Hence for the proposition, it is sufficient to show that, at all points in $S$,
    \begin{equation*}
        R_{abcd}=\bm{0}\Leftrightarrow\mathcal{R}\indices{^a_b_c_d}=\bm{0},
    \end{equation*}
    where $\mathcal{R}\indices{^a_b_c_d}=\bm{0}$ is the Riemann tensor on $S$ associated with $D$. {\color{teal} ($\mathcal{R}\indices{^a_b_c_d}$ is defined such that, for all smooth fields $v^b$ on $S$ spacelike relative to $S$, $\mathcal{R}\indices{^a_b_c_d} v^b := - 2 D_{[c} D_{d]} v^a$.)}
    For this, note that the right hand side is equivalent to the requirement that, for any spacelike covector field $\sigma_a$ on $S$,
    \begin{equation*}
        \bm{0}=\mathcal{R}\indices{^n_a_b_c}\sigma_n=2D_{[b}D_{c]}\sigma_a=2\hat{h}\indices{^n_b}\hat{h}\indices{^m_c}\nabla_{[n}\nabla_{m]}\sigma_a=\hat{h}\indices{^n_b}\hat{h}\indices{^m_c}R\indices{^r_a_n_m}\sigma_r.
    \end{equation*}
    Hence, it is equivalent to the requirement that 
    \begin{align*}
        \bm{0}=h_{ar}\hat{h}\indices{^n_c}\hat{h}\indices{^m_d}R\indices{^r_b_n_m}=\hat{h}\indices{^n_c}\hat{h}\indices{^m_d}R\indices{_a_b_n_m}=(\delta\indices{^n_c}-t_c\xi^n)(\delta\indices{^m_d}-t_d\xi^m)R\indices{_a_b_n_m}=R\indices{_a_b_c_d},
    \end{align*}
    where we have made use of \eqref{eq:downstairsnewtonian} and \eqref{eq:riemannantisymmetric}.
\end{proof}
(One might be tempted to say that a Carroll spacetime $\langle M, h_{ab}, \xi^a, \nabla\rangle$ is spatially flat iff for any unit timelike covector field $t_a$ and any spacelike hypersurface $S$ relative to $t_a$, parallel transport of spacelike \textit{vectors} within $S$ is (at least locally) path-independent. But this will not work, since any Carroll spacetime which satisfied this condition would have to be flat simpliciter, i.e.~$R\indices{^a_b_c_d}=\bm{0}$.)

The following proposition gives a characterisation of the relative strengths of the curvature conditions $R\indices{^a_b_c_d}=\bm{0}$ and $R_{abcd}=\bm{0}$:
\begin{prop}\label{prop:theversionof4.2.4}
    Let $\langle M, h_{ab}, \xi^a, \nabla\rangle$ be a Carroll spacetime which is spatially flat ($R_{abcd}=\bm{0}$). Then $R\indices{^a_b_c_d}=\bm{0}$ iff (at least locally) there exists a unit timelike covector field $t_a$ such that $\nabla_at_b=\bm{0}$.
\end{prop}
\begin{proof}
    For the `if' direction, suppose that such a unit timelike covector field exists. Since $R_{abcd}=\bm{0}$, it suffices to show that $t_nR\indices{^n_b_c_d}=\bm{0}$. But
    $t_nR\indices{^n_b_c_d}=2\nabla_{[c}\nabla_{d]}t_b=\bm{0}$.
    For the `only if' direction, let $p\in M$ and let $\overset{p}{t}_a$ be an arbitrary unit timelike covector at $p$, and let $O$ be some open set containing $p$. Let $t_a$ be the covector field which results from parallel transporting $\overset{p}{t}_a$, along any curve(s), throughout $O$ (this makes sense, since $R\indices{^a_b_c_d}=\bm{0}$ implies that the resulting covector field will be independent of the choice). By construction, $t_a$ is unit timelike, and $\nabla_at_b=\bm{0}$.
\end{proof}
To understand the intuitive geometric significance of proposition \ref{prop:theversionof4.2.4}, note that if $t_a$ is a unit timelike covector field such that $\nabla_at_b=\bm{0}$, then for any spacelike hypersurface $S$ relative to $t_a$, $D$ and $\nabla$ induce the same conditions for parallel transport of spacelike vectors relative to $t_a$ in $S$. Thus since $R_{abcd}=\bm{0}$, $\mathcal{R}\indices{^a_b_c_d}=\bm{0}$, and parallel transport of spacelike vectors relative to $t_a$ within any such $S$ is path-independent. Finally, take any three mutually orthogonal vectors which are spacelike relative to $t_a$ at some $p\in M$, and parallel transport them (along any curve(s)) throughout the spacelike hypersurface $S$ relative to $t_a$ containing $p$. Then parallel transport them throughout $M$ along $\xi^a$. The resulting vector fields are constant, and, together with $\xi^a$, form a basis for the tangent space at each point.

The final curvature condition we will consider is $\xi^nR\indices{^a_b_c_n}= \bm{0}$. We have the following proposition:
\begin{prop}\label{prop:funcommutingdiagramwithoutthediagram}
    Let $\langle M, h_{ab}, \xi^a, \nabla\rangle$ be a Carroll spacetime. Then the following conditions are equivalent:
    \begin{itemize}
        \item[(i)] $\xi^nR\indices{^a_b_c_n}= \bm{0}$.
        \item[(ii)] The operator $\pounds_\xi$ commutes with $\nabla$ in its action on all smooth tensor fields.
        \item[(iii)] There exists (at least locally) a unit timelike covector field $t_a$ such that $\pounds_\xi t_a=\bm{0}$ and $\pounds_\xi \nabla_at_b=\bm{0}$.
    \end{itemize}
    Moreover, if the above conditions hold, then $\pounds_\xi R\indices{^a_b_c_d}=\bm{0}$.  
\end{prop}
\begin{proof}
    The equivalence of (i) and (ii) follows from the compatibility conditions and problem 1.8.3 of \parencite{Malament2012-MALTIT}. That (ii) implies $\pounds_\xi R\indices{^a_b_c_d}=\bm{0}$ follows from  problem 1.9.4 of \parencite{Malament2012-MALTIT}. For the equivalence of (i) and (iii), note that for any unit timelike covector field $t_a$, we have
    \begin{align}
        t_m\xi^nR\indices{^m_b_c_n}&=2\xi^n\nabla_{[c}\nabla_{n]}t_b\nonumber\\
        &=\nabla_{c}\xi^n\nabla_{n}t_b-\xi^n\nabla_{n}\nabla_{c}t_b.\label{eq:unittimelikexi}
    \end{align}
    Thus since $\xi^nR\indices{_a_b_c_n}=\bm{0}$, if there exists a unit timelike covector field such that $\pounds_\xi t_a=\bm{0}$ and $\pounds_\xi \nabla_at_b=\bm{0}$, then $\xi^nR\indices{^a_b_c_n}= \bm{0}$. Conversely, if $\xi^nR\indices{^a_b_c_n}=\bm{0}$, then by \eqref{eq:unittimelikexi} we must have that $\pounds_\xi \nabla_at_b=\bm{0}$ for \textit{any} unit timelike covector field such that $\pounds_\xi t_a=\bm{0}$. But such a unit timelike covector field always exists (at least locally): consider the restriction of any unit timelike covector field to some spacelike hypersurface $S$ and then parallel transport it throughout some open region $O$ along $\xi^a$. So we have (iii).
\end{proof}
Proposition \ref{prop:funcommutingdiagramwithoutthediagram} provides the following interpretation of the curvature condition $\xi^nR\indices{^a_b_c_n}= \bm{0}$. The conditions (i)--(iii) along with the compatibility conditions tell us that if $\xi^nR\indices{^a_b_c_n}= \bm{0}$, then $R\indices{^a_b_c_d}$ is constant along the integral curves of $\xi^a$. When this condition holds, then facts about the geometry of any spacelike hypersurface completely determine the geometry of the entire Carroll spacetime. On the one hand, this property is absolutely crucial if one wishes to e.g.\ make use of the properties of spacelike geodesics at a point to characterise the Ricci curvature there (as we shall see in the following proposition), since the timelike geodesics of a Carroll spacetime do not give us any information about the degrees of freedom of the connection which are not fixed by the compatibility conditions. On the other hand, one sees that Carroll spacetime structure does not automatically preclude the possibility of non-trivial `temporal evolution' for spacetimes where $\xi^nR\indices{^a_b_c_n}\neq\bm{0}$ (in the sense that $R\indices{^a_b_c_d}$ is not automatically fixed uniquely by its value on any spacelike hypersurface).

Having discussed in some detail the geometrical structure of Carroll spacetimes, we now move on to consider dynamics. We will assume that the stress-energy content is represented by a symmetric tensor field $T^{ab}$ (intuitively, if $t_a$ is a unit timelike vector field, then $t_nT^{na}$ represents the four-momentum relative to spacelike hypersurfaces relative to $t_a$). Let $T:=h_{nm}T^{nm}$. Dynamics are then given by:
\begin{align}
    R_{ab}&=8\pi(T_{ab}-1/2h_{ab}T),\label{eq:carrollefes}\\
    \nabla_nT^{na}&=\bm{0},\label{eq:covconservation}
\end{align}
where we have lowered indices with $h_{ab}$. {\color{teal} (These equations of motion can be obtained by varying the magnetic Carroll gravity action, on which see \textcite[eq.\ 1.4]{Henneaux2}. \color{blue} Note that in order to obtain our \eqref{eq:carrollefes} and \eqref{eq:covconservation} by varying the action for magnetic Carroll gravity given by \textcite{Henneaux2}, one either (to use the notation of that article) needs to set $\pi^{ij} \equiv 0$ or needs to impose some condition on the stress energy content; our thanks to an anonymous reviewer for inviting us to mention this latter point.)} The following proposition gives an interpretation of \eqref{eq:carrollefes}:
\begin{prop}
    Let $\langle M, \xi^a, h_{ab}, \nabla\rangle$ be a Carroll spacetime satisfying $\xi^nR\indices{^a_b_c_n}=\bm{0}$, and let $T^{ab}$ be a smooth symmetric field on $M$. Then for all points $p\in M$, \eqref{eq:carrollefes} holds at $p$ iff for any unit spacelike vector field $\sigma^a$ which is geodesic with respect to $\nabla$, the average spatial relative acceleration (ASRA) of $\sigma^a$ at $p$ satisfies
    \begin{equation}\label{eq:ASRA}
        ASRA:=-\frac{1}{2}\sum_{i=1}^{2}\overset{i}{\lambda}_r\sigma^n\nabla_n(\sigma^m\nabla_m\overset{i}{\lambda}\vphantom{\lambda}^r)=4\pi(T_{nm}-1/2h_{nm}T)\sigma^n\sigma^m ,
    \end{equation}
    where $\overset{i}{\lambda}\vphantom{\lambda}^a$, $i=1,2$ are any two connecting fields for $\sigma^a$ which are spacelike and mutually orthonormal at $p$.
\end{prop}  
\begin{proof}
    First, note that 
    \begin{align*}
        ASRA&=-\frac{1}{2}\sum_{i=1}^{2}\overset{i}{\lambda}_r\sigma^n\nabla_n(\sigma^m\nabla_m\overset{i}{\lambda}\vphantom{\lambda}^r)\\
        &=-\frac{1}{2}\sum_{i=1}^{2}\overset{i}{\lambda}_rR\indices{^r_n_m_s}\sigma^n\overset{i}{\lambda}\vphantom{\lambda}^m\sigma^s\\
        &=-\frac{1}{2}R\indices{^r_n_m_s}\sigma^n\sigma^s\sum_{i=1}^{2}\overset{i}{\lambda}_r\overset{i}{\lambda}\vphantom{\lambda}^m.
    \end{align*}
    The orthonormality condition implies that, at $p$
    \begin{equation*}
        \sum_{i=1}^{2}\overset{i}{\lambda}_b\overset{i}{\lambda}\vphantom{\lambda}^a=\delta\indices{^a_b}-\sigma^a\sigma_b-t_b\xi^a
    \end{equation*}
    where $t_a$ is any unit timelike covector field which annihilates $\sigma^a$ and the $\overset{i}{\lambda}\vphantom{\lambda}^a$ at $p$ (contraction on both sides with $\sigma^a$, $\xi^a$, and the $\overset{i}{\lambda}\vphantom{\lambda}^a$ yields the same result). Note that such a unit timelike covector field always exists. Thus
    \begin{align*}
        ASRA=-\frac{1}{2}R\indices{^r_n_m_s}\sigma^n\sigma^s(\delta\indices{^m_r}-\sigma^m\sigma_r-t_r\xi^m).
    \end{align*}
    The second term on the right hand side vanishes by \eqref{eq:riemannantisymmetric}. The third term on the right hand side also vanishes, by \eqref{eq:riemannantisymmetric} and using that $\xi^nR\indices{^a_b_c_n}=\bm{0}$. So we have
    \begin{align*}
        ASRA&=-\frac{1}{2}\delta\indices{^m_r}R\indices{^r_n_m_s}\sigma^n\sigma^s\\
        &=\frac{1}{2}\delta\indices{^m_r}(R\indices{^r_m_s_n}+R\indices{^r_s_n_m})\sigma^n\sigma^s\\
        &=\frac{1}{2}\delta\indices{^m_r}R\indices{^r_s_n_m}\sigma^n\sigma^s\\
        &=\frac{1}{2}R\indices{_n_m}\sigma^n\sigma^m ,
    \end{align*}
    where we have made use of \eqref{eq:bianchi1} in the second equality and \eqref{eq:riemannantisymmetric} in the third. Thus, if \eqref{eq:carrollefes} holds, then $ASRA=4\pi(T_{nm}-1/2h_{nm}T)\sigma^n\sigma^m$. Conversely, given any unit spacelike vector $\overset{p}{\sigma}\vphantom{\sigma}^a$ at some $p\in M$, we can always find a spacelike vector field $\sigma^a$ extending $\overset{p}{\sigma}\vphantom{\sigma}^a$ which is geodesic. Thus $R_{nm}\overset{p}{\sigma}\vphantom{\sigma}^n\overset{p}{\sigma}\vphantom{\sigma}^m=8\pi(T_{nm}-1/2h_{nm}T)\overset{p}{\sigma}\vphantom{\sigma}^n\overset{p}{\sigma}\vphantom{\sigma}^m$ for any unit spacelike vector at $p$. But contracting both sides of \eqref{eq:carrollefes} with $\xi^a\xi^b$ or $\xi^a\overset{p}{\sigma}\vphantom{\sigma}^b$ yields zero. So \eqref{eq:carrollefes} must hold at $p$.
    
\end{proof}

This result is analogous to \parencite[Proposition 2.7.2]{Malament2012-MALTIT}, in which it is shown that the obtaining of Einstein's equations is equivalent to the `average radial acceleration' for all geodesic reference frames taking a specific form, analogous to the RHS of \eqref{eq:ASRA} but in the relativistic setting. In both cases, the point is that when the dynamical equations of the theory obtain, the content of those equations is very directly encoded in terms of facts about average (spatial) radial acceleration.

As in general relativity, there are a variety of `energy conditions' one can impose on $T^{ab}$ in Carroll spacetimes. {\color{teal} (For energy conditions in general relativity, see \textcite{Curiel2016-CURAPO}.)} One option is the following:
\begin{description}
    \item[Strengthened dominant Carroll energy condition:] For any $p\in M$ and any timelike covector $\mu_a$ at $p$, $\mu_n\mu_mT^{nm}\geq 0$ and either $T^{ab}=\bm{0}$ or $\mu_nT^{na}$ is timelike. 
\end{description}
The strengthened dominant Carroll energy condition implies that $T^{ab}=\rho\xi^a\xi^b$, where $\rho:=t_nt_mT^{nm}$ for any unit timelike covector field $t_a$. In this case, \eqref{eq:carrollefes} and \eqref{eq:covconservation} become 
\begin{align}
    R_{ab}&=\bm{0},\\
    \xi^n\nabla_n\rho&=0.
\end{align}
In other words, the matter dynamics become trivial: fluid elements necessarily traverse the integral curves of $\xi^a$ and the energy density $\rho$ is constant along any such integral curve. However, there are also weaker energy conditions available, for example:
\begin{description}
    \item[Weak Carroll energy condition:] For any $p\in M$ and any timelike covector $\mu_a$ at $p$, $\mu_n\mu_mT^{nm}\geq 0$. 
\end{description}
The weak Carroll energy condition is compatible with non-trivial matter dynamics; however, as a consequence, it also allows for stress-energy propogation along spacelike (as well as timelike) curves. We discuss this further in \S\ref{sec:discussion}.

\section{The ultra-relativistic limit}\label{sec:limit}

Having presented the basic structure of Carroll gravity, we'll now consider the sense in which this theory is the ultra-relativistic limit of general relativity. While of course there is already a literature on this topic---see in particular \parencite{Lévy1965, Dautcourt1998OnTU, Duval_2014, HansenPaper, HansenThesis}---in this article we'll take a more `geometrical' approach to the limit (\S\ref{sec:limitmath} and \S\ref{sec:frame}), before comparing with those existing approaches to the ultra-relativistic limit (\S\ref{sec:comparison}).

\subsection{A `geometric' approach to the limit}\label{sec:limitmath}

Our approach will mirror that of \textcite{1986_Malament} for the non-relativistic limit. Let $g_{ab}(\lambda)$ be a one-parameter family of (non-degenerate) Lorentzian metrics on $M$, where $\lambda\in[0,k]$ for some $k$. For the ultra-relativistic limit, we are interested in the case where $g_{ab}(\lambda)$ satisfies two conditions:
\begin{enumerate}
    \item $\lambda g_{ab}(\lambda)\rightarrow - h_{ab}$ as $\lambda\rightarrow 0$ for some field $h_{ab}$ of signature $(0,1,1,1)$.
    \item $g^{ab}(\lambda)\rightarrow \xi^a\xi^b$ as $\lambda\rightarrow 0$ for some non-zero vector field $\xi^a$ such that $\pounds_\xi h_{ab}=\bm{0}$.
\end{enumerate}
Here, $\lambda$ corresponds to $c^2$; changing the value of $\lambda$ then amounts to narrowing the lightcones, and so provides the resources to take, in a `geometrical' way, an ultra-relativistic limit.
Geometrically, one can understand this as the lightcones narrowing until they become a congruence of curves (a `fibration'); i.e.~the integral curves of the vector field $\xi^a$. {\color{blue} (Note that: (i) the scaling approach considered here is in principle not limited to the torsion-free case; (ii) the defining conditions for $g_{ab}(\lambda)$ are precisely satisfied based upon the scaling relations for the ADM components of the metric given in \textcite[eq.\ 13]{Niedermaier}.)}

\begin{prop}\label{prop:carrollianlimits1}
    Let $g_{ab}(\lambda)$ be a one-parameter family of Lorentzian metrics on $M$, and for each $g_{ab}(\lambda)$ let $\overset{\lambda}{\nabla}$ be the associated Levi-Civita derivative operator. Let $\xi^a$, $h_{ab}$ be as in conditions (1) and (2). Then:
    \begin{itemize}
        \item[(i)] There is a derivative operator $\nabla$ on $M$ satisfying $\overset{\lambda}{\nabla}\rightarrow \nabla$ as $\lambda\rightarrow 0$.
        \item[(ii)] $\langle M, h_{ab}, \xi^a, \nabla\rangle$ is a Carroll spacetime.
    \end{itemize}
\end{prop}
\begin{proof}
    Since $\lambda g_{ab}(\lambda)\rightarrow -h_{ab}$ smoothly, there must exist fields $v_{ab}$, $s_{ab}(\lambda)$, $s_{ab}$ satisfying 
    \begin{equation}\label{eq:spatialexpansion}
        \lambda g_{ab}(\lambda)=-h_{ab}+\lambda v_{ab}+\lambda^2s_{ab}(\lambda)
    \end{equation}
    \begin{equation*}
        s_{ab}(\lambda)\rightarrow s_{ab}\; \text{as}\; \lambda\rightarrow 0.
    \end{equation*}
    (i.e.~the limit $\lambda\rightarrow 0$ is twice-differentiable). Similarly, there must exist fields $h^{ab}$, $p^{ab}(\lambda)$, $p^{ab}$ satisfying 
    \begin{equation}\label{eq:temporalexpansion}
        g^{ab}(\lambda)=\xi^a\xi^b-\lambda h^{ab}+\lambda^2p^{ab}(\lambda)
    \end{equation}
    \begin{equation*}
        p^{ab}(\lambda)\rightarrow p^{ab}\; \text{as}\; \lambda\rightarrow 0.
    \end{equation*}
Since $-\lambda\delta\indices{^a_b} =\lambda g_{bn}(\lambda)g^{na}(\lambda)$, we have
\begin{equation*}
    -\lambda\delta\indices{^a_b}=-h_{bn}\xi^n\xi^a+\lambda(v_{an}\xi^n\xi^a+h_{an}h^{na})+\lambda^2(...),
\end{equation*}
so that in the limit $\lambda\rightarrow 0$, 
\begin{equation}\label{eq:compatibility}
    h_{an}\xi^n= \bm{0}.
\end{equation}
Next, let $\Tilde{\nabla}$ be an arbitrary derivative operator on $M$ (such always exist locally). We know that $\overset{\lambda}{\nabla}=(\Tilde{\nabla}, C\indices{^a_b_c}(\lambda))$, where
\begin{equation*}
    C\indices{^a_b_c}(\lambda)=1/2g^{an}(\lambda)(\Tilde{\nabla}_ng_{bc}(\lambda)-\Tilde{\nabla}_bg_{nc}(\lambda)-\Tilde{\nabla}_cg_{nb}(\lambda)).
\end{equation*}
Using \eqref{eq:spatialexpansion} and \eqref{eq:temporalexpansion}, we have
\begin{align*}
    C\indices{^a_b_c}(\lambda)=1/2(\xi^a\xi^n&-\lambda h^{an}+\lambda^2p^{an}(\lambda))(-\lambda^{-1}[\Tilde{\nabla}_nh_{bc}-\Tilde{\nabla}_bh_{nc}-\Tilde{\nabla}_ch_{nb}] \\ &\qquad + V_{nbc}+\lambda S_{nbc}(\lambda)),
\end{align*}
where \[V_{abc}=\Tilde{\nabla}_av_{bc}-\Tilde{\nabla}_bv_{ac}-\Tilde{\nabla}_cv_{ab}\]
and \[S_{abc}(\lambda)=\Tilde{\nabla}_as_{bc}(\lambda)-\Tilde{\nabla}_bs_{ac}(\lambda)-\Tilde{\nabla}_cs_{ab}(\lambda).\] But we know that 
\begin{align*}
    \pounds_\xi h_{ab}&=\xi^n\Tilde{\nabla}_nh_{ab}+h_{nb}\Tilde{\nabla}_a\xi^n+h_{an}\Tilde{\nabla}_b\xi^n\\
    &=\xi^n\Tilde{\nabla}_nh_{ab}-\xi^n\Tilde{\nabla}_ah_{nb}-\xi^n\Tilde{\nabla}_bh_{an}\\
    &=\bm{0}
\end{align*}
where we have made use of \eqref{eq:compatibility}, so that 
\begin{align*}
    C\indices{^a_b_c}(\lambda)=-1/2(- h^{an}&+\lambda p^{an}(\lambda))(\Tilde{\nabla}_nh_{bc}-\Tilde{\nabla}_bh_{nc}-\Tilde{\nabla}_ch_{nb})\\
    &+1/2(\xi^a\xi^n-\lambda h^{an}+\lambda^2p^{an}(\lambda))(V_{nbc}+\lambda S_{nbc}(\lambda)).
\end{align*}
Hence, if we define \[C\indices{^a_b_c}=1/2h^{an}(\Tilde{\nabla}_nh_{bc}-\Tilde{\nabla}_bh_{nc}-\Tilde{\nabla}_ch_{nb})+1/2\xi^a\xi^nV_{nbc},\] then it follows that $C\indices{^a_b_c}(\lambda)\rightarrow C\indices{^a_b_c}$ as $\lambda\rightarrow 0$ and hence that $\overset{\lambda}{\nabla}\rightarrow \nabla=(\Tilde{\nabla}, C\indices{^a_b_c})$. Finally we then have that
\begin{equation*}
    \overset{\lambda}{\nabla}_ag^{bc}(\lambda) \rightarrow\nabla_a(\xi^b\xi^c)\; \text{as} \; \lambda\rightarrow 0
\end{equation*}
and
\begin{equation*}
    \overset{\lambda}{\nabla}_a\lambda g_{bc}(\lambda) \rightarrow\nabla_ah_{bc}\; \text{as} \; \lambda\rightarrow 0
\end{equation*}
so that $\nabla_a\xi^b=\bm{0}$ and $\nabla_ah_{bc}=\bm{0}$.
\end{proof}

So, in the lightcone-narrowing limit $\lambda \rightarrow 0$, a Lorentzian spacetime will converge to a Carroll spacetime. One also sees that the condition $\pounds_\xi h_{ab}=\bm{0}$ is necessary for convergence in the limit. But what of dynamics? Suppose now in addition that for each $\lambda$ we also have a (symmetric) stress-energy tensor $T_{ab}$, and suppose that
\begin{itemize}
    \item[3.] Einstein's equation $\overset{\lambda}{R}_{ab}=8\pi(T_{ab}(\lambda)-1/2g_{ab}(\lambda)T(\lambda))$ holds for all $\lambda$.
    \item[4.] $\lambda^{-2} T^{ab}(\lambda)\rightarrow T^{ab}$ as $\lambda\rightarrow 0$ for some $T^{ab}$.
\end{itemize}
where $T(\lambda):=T_{nm}(\lambda)g^{nm}(\lambda)$. 

\begin{prop}\label{prop:carrollianlimits2}
Let $g_{ab}(\lambda)$ be a one-parameter family of metrics on $M$ which, together with the symmetric family $T^{ab}(\lambda)$, satisfies conditions (1)--(4). Let $\langle M, h_{ab}, \xi^a, \nabla\rangle$ be the limit Carroll spacetime obtained in proposition \ref{prop:carrollianlimits1}. Then there exist a symmetric field $T_{ab}$ and a field $T$ on $M$ satisfying
\begin{itemize}
    \item[(i)] $T_{ab}(\lambda)\rightarrow T_{ab}$ as $\lambda\rightarrow0$.
    \item[(ii)] $\lambda^{-1}T(\lambda)\rightarrow -T$ as $\lambda\rightarrow0$.
    \item[(iii)] $R_{ab}=8\pi(T_{ab}-1/2h_{ab}T)$.
    \item[(iv)] $\nabla_nT^{na}=\bm{0}$. 
\end{itemize}   
\end{prop}
\begin{proof}
    Since $\lambda^{-2} T^{ab}(\lambda)\rightarrow T^{ab}$ smoothly there must exist fields $t^{ab}$, $u^{ab}(\lambda)$, $u^{ab}$ satisfying 
\begin{equation}\label{eq:stressenergyexpansion}
        \lambda^{-2} T^{ab}(\lambda)=T^{ab}+\lambda t^{ab}+\lambda^2u^{ab}(\lambda),
    \end{equation}
    with
    \begin{equation*}
        u^{ab}(\lambda)\rightarrow u^{ab}\; \text{as}\; \lambda\rightarrow 0.
    \end{equation*}
    Thus, from \eqref{eq:spatialexpansion} and \eqref{eq:stressenergyexpansion}: 
    \begin{align*}
        T_{ab}(\lambda)&=g_{an}(\lambda)g_{bm}(\lambda)T^{nm}(\lambda)\\        
        &=(-\lambda^{-1} h_{an}+v_{an}+\lambda s_{an}(\lambda))(-\lambda^{-1} h_{bm}+v_{bm}+\lambda s_{bm}(\lambda)) \\ &\qquad \times (\lambda^2T^{nm}+\lambda ^3t^{nm}+\lambda^4u^{nm}(\lambda)),
    \end{align*}
    so that defining $T_{ab}=g_{an}g_{bm}T^{nm}$ we have $T_{ab}(\lambda)\rightarrow T_{ab}$ as $\lambda\rightarrow0$. Similarly:
    \begin{align*}
        \lambda^{-1}T(\lambda)&=\lambda g_{nm}(\lambda)\lambda^{-2}T^{nm}(\lambda)\\
        &=(-h_{nm}+\lambda v_{nm}+\lambda^2 s_{nm}(\lambda))(T^{nm}+\lambda t^{nm}+\lambda^2u^{nm}(\lambda)),
    \end{align*}
    so that defining $T=h_{nm}T^{nm}$, $\lambda^{-1} T(\lambda)\rightarrow -T$ as $\lambda\rightarrow0$. Finally, 
    \begin{align*}
        g_{ab}(\lambda)T(\lambda)&=\lambda g_{ab}\lambda g_{nm}(\lambda)\lambda^{-2}T^{nm}(\lambda)\\
        &=(-h_{ab}+\lambda v_{ab}+\lambda^2 s_{ab}(\lambda))(-h_{nm}+\lambda v_{nm}+\lambda^2 s_{nm}(\lambda))\\ &\qquad \times (T^{nm}+\lambda t^{nm}+\lambda^2u^{nm}(\lambda)).
    \end{align*}
    Since $\overset{\lambda}{R}_{ab}\rightarrow R_{ab}$ as $\lambda\rightarrow 0$, it follows that $R_{ab}=8\pi(T_{ab}-1/2h_{ab}T)$. And since 
    \begin{equation*}
        \overset{\lambda}{\nabla}_n\lambda^{-2}T^{na}(\lambda)\rightarrow \nabla_nT^{na}\;\text{as}\;\lambda\rightarrow 0
    \end{equation*}
    and (3) implies that $\overset{\lambda}{\nabla}_nT^{na}(\lambda)=0$ for all $\lambda$, we have $\nabla_nT^{na}=\bm{0}$.
\end{proof}

So, we have taken a `geometrical', lightcone-narrowing limit of general relativity, and have obtained Carroll gravity. We are far from the first to consider the ultra-relativistic limit of general relativity, and in \S\ref{sec:comparison} we will compare our approach with what has come before (in particular with the work of \textcite{Dautcourt1998OnTU, HansenPaper, HansenThesis}). Before doing so, however, some words on how the foregoing work can be set in the broader context of the `frame theory' developed by \textcite{Ehlers}.

\subsection{Understanding the limit via frame theory}\label{sec:frame}

Consider again the non-relativistic, lightcone-widening limit of general relativity, as presented by \textcite{1986_Malament}. In order to make such a limit completely rigorous, it is convenient to avail oneself of the resources of `frame theory': a unified spacetime framework (hence the name) developed by \textcite{Ehlers} which encompasses both relativistic and non-relativistic spacetime models. One can then---\emph{à la} \textcite{Fletcher2019-FLEOTR-3}---make rigorous sense of the non-relativistic limit by imposing a topology on this space of models and then considering a limit in this topology.

In more detail (and following the presentation by \textcite{Fletcher2019-FLEOTR-3}): the models of frame theory are $\langle M , t_{ab} , s^{ab}, \nabla, T^{ab} \rangle$, where $t_{ab}$ and $s^{ab}$ are (respectively) symmetric temporal and spatial metrics, $\nabla$ is a torsion-free derivative operator compatible with $t_{ab}$ and $s^{ab}$, and $T^{ab}$ is the stress-energy tensor. The key generalisation offered by frame theory is that, at this point, one does not specify the signatures of $t_{ab}$ and $s^{ab}$; moreover, one does not impose orthogonality, but only the weaker condition that $t_{ab} s^{bc} = \kappa \delta\indices{^c_a}$ (where $\kappa$ is known as the `causality constant' of the model); as a result, the theory encompasses both Lorentizan spacetimes (where $t_{ab} = g_{ab}$ and $s^{ab} = - \kappa g^{ab}$ and $\kappa = c^{-2}$) and non-relativistic spacetimes (where $s^{ab} = h^{ab}$ and $\kappa = 0$).

Frame theory doesn't purport to offer a deep \emph{physical} sense of unification of relativistic and non-relativistic spacetime theories, at least if unification is understood along the lines presented by e.g.\ \textcite{Maudlin1996-MAUOTU-2}. However, as already pointed out, it does afford the resources to make rigorous sense of a geometrical non-relativistic limit. For example, helping oneself to the $C^2$ point-open product topology on the space of models of frame theory (see \textcite[\S3]{Fletcher2019-FLEOTR-3}), one can define the non-relativistic limit of a family of relativistic spacetimes as follows:
\begin{definition}
    \textbf{\em (Newtonian limit, Ehlers)} Let $\langle M , \overset{\lambda}{t}_{ab}, \overset{\lambda}{s}{}^{ab}, \overset{\lambda}{\nabla} , \overset{\lambda}{T}{}^{ab} \rangle$ with $\lambda \in ( 0, a)$ for some $a>0$ be a one-parameter family of models of general relativity. Then $\langle M , t_{ab}, s^{ab}, \nabla , T^{ab} \rangle$ is a `Newtonian limit' of the family when it is a model of Newton-Cartan theory and
    \[ \lim_{\lambda \rightarrow 0} \left( \overset{\lambda}{t}_{ab}, \overset{\lambda}{s}{}^{ab}, \overset{\lambda}{\nabla} , \overset{\lambda}{T}{}^{ab} \right) =  \left( t_{ab}, s^{ab}, \nabla , T^{ab} \right) \] in the $C^2$ point-open product topology.
\end{definition}
\noindent ({\color{teal} This definition is due to Ehlers, but we use the terminology of \textcite{Fletcher2019-FLEOTR-3}. Note that defining here the $C^k$ point-open product topology would be a little involved; for explicit details, we refer the reader to \textcite{Fletcher2019-FLEOTR-3}.} \textcite{Fletcher2019-FLEOTR-3} {\color{blue}also} discusses other possible choices of topology on the space of models of frame theory; we won't go into this in further detail in this article.)

The main observation which we wish to make here is that frame theory is already equipped to encompass the models of Carroll gravity (an so is more than the mere union of models of general relativity and models of non-relativistic gravity): one simply identifies $s^{ab} = \xi^a \xi^b$ and $t_{ab} = h_{ab}$ and imposes (as in the Newton-Cartan case) that $\kappa = 0$. Then, one can just as well write down the following definition of an ultra-relativistic limit in frame theory:
\begin{definition}
    \textbf{\em (Ultra-relativistic limit)} Let $\langle M , \overset{\lambda}{t}_{ab}, \overset{\lambda}{s}{}^{ab}, \overset{\lambda}{\nabla} , \overset{\lambda}{T}{}^{ab} \rangle$ with $\lambda \in ( 0, a)$ for some $a>0$ be a one-parameter family of models of general relativity. Then $\langle M , t_{ab}, s^{ab}, \nabla , T^{ab} \rangle$ is an `ultra-relativistic limit' of the family when it is a model of Carroll gravity and
    \[ \lim_{\lambda \rightarrow 0} \left( \overset{\lambda}{t}_{ab}, \overset{\lambda}{s}{}^{ab}, \overset{\lambda}{\nabla} , \overset{\lambda}{T}{}^{ab} \right) =  \left( h_{ab}, s^{ab}, \nabla , T^{ab} \right) \] in the $C^2$ point-open product topology.
\end{definition}
\noindent (Of course, as before, other topologies on the space of models are available; note also that this $\lambda$ is clearly different from that in the previous definition.)
So---to repeat---this geometrical approach to the ultra-relativistic limit can be made rigorous using frame theory. Of course, this leaves open a more thoroughgoing investigation of ultra-relativistic limits of particular relativistic solutions within frame theory (i.e.,\ an investigation analogous to that undertaken by \textcite{Fletcher2019-FLEOTR-3} for the non-relativistic limit); this, however, would take us too far from our ambitions for this article, and so will have to wait for another day.

\subsection{Comparison with other approaches to the limit}\label{sec:comparison}

In the case of the non-relativistic limit of general relativity, it's by now acknowledged that taking a geometrical approach to the limit is not \emph{per se} incompatible with approaches to the limit in terms of series expansions (in say $1/c$){\color{blue}---this in fact should be obvious, given that $\lambda$-scaling leads to a series expansion in $\lambda$ (although one merit of the $\lambda$-scaling approach is that it is more geometrically `intrinsic')}. {\color{teal} (This point is made by \textcite{Fletcher2019-FLEOTR-3, Hartong2022ReviewON}.)} In this subsection, we'll make an analogous point---namely, that our `geometrical' approach to the ultra-relativistic limit isn't incompatible with other approaches to the limit in terms of series expansions. {\color{teal} (It's also not incompatible with approaches to obtaining Carroll gravity via dimensional reduction---we won't discuss those approaches further here, but see \parencite{Duval_2014, HansenThesis} for further details.)}

There is a straightforward point to be made here, alongside a deeper point. The straightforward point is this. Assuming that objects such as $\overset{\lambda}{t}_{ab}$ are functions of some $\lambda$ and have finite limits for $\lambda \rightarrow 0$ should already imply that one can consider a Taylor expansion around $\lambda = 0$---hence, no incompatibility. {\color{blue} (This is essentially the same point as made above, only now in the context of the ultra-relativistic limit.)}

The deeper point is this. The approach to both the non-relativistic limit and ultra-relativistic limit developed in recent works such as \parencite{Hartong2022ReviewON, HansenThesis} is in fact somewhat novel compared with the work on limits of general relativistic spacetime models which preceded it. {\color{teal} (It is this novelty which, in the case of the non-relativistic limit, has allowed these authors to construct a new, `Type II' version of Newton--Cartan theory. This theory has various interesting features---for example, as pointed out by \textcite{Obers}, it admits an action principle. For further philosophical discussion of Type II Newton--Cartan theory, see \parencite{WolfReadSanchioni}.)} This novelty lies in the fact that these authors do not begin with the standard objects of general relativity when taking the limit. Rather, in the non-relativistic case, they begin with a more general connection built from `pre-non-relativistic' variables (see \textcite[\S6.1]{Hartong2022ReviewON}; the approach goes back to \textcite{VDB})---a connection which generically has torsion!---and then perform a series expansion of that connection. (Hence, it should be of little surprise that Type II Newton--Cartan theory is compatible with spacetime torsion.) In the case of the ultra-relativistic limit, \textcite{HansenPaper, HansenThesis} have followed a similar approach, building a connection in terms of `pre-ultra-relativistic' variables; again, the result they obtain has the potential to be more general than what we have obtained here, but \emph{ipso facto} isn't incompatible with our results. (Recall also that, as already mentioned in \S\ref{sec:intro}, these approaches are related to the `electric' version of Carroll gravity which we do not consider in this article.)

\section{Conceptual assessment}\label{sec:discussion}

Having presented (magnetic) Carroll gravity and clarified the sense in which it is an ultra-relativistic limit of general relativity, we now discuss some of its physically interesting properties.



Widening the light cones of a general relativistic spacetime gives rise to a non-relativistic spacetime, in which it is well-known that (i) for any two spacetime points, there is a well-defined absolute temporal distance between them, but (ii) it's not the case that for any two spacetime points, there is a well-defined absolute spatial distance between them (rather, this is true only for co-temporal spacetime points). The situation is reversed in a Carrollian spacetime: (i$'$) for any two spacetime points, there is a well-defined absolute \emph{spatial} distance between them, but (ii$'$) it's not the case that for any two spacetime points, there is a well-defined \emph{temporal} distance between them (rather, this is true only for co-spatial spacetime points). This role-reversal is one manifestation of a broader `duality' between non-relativistic and Carrollian spacetime structures---for more discussion of which, see \parencite{Duval_2014}.

Let's {\color{blue}home} in on two particularly interesting features of Carroll spacetimes which were identified by \textcite{Lévy1965} in the passage quoted in \S\ref{sec:intro}:
\begin{enumerate}
    \item An attenuated notion of causality, in the sense that almost every pair of events in spacetime is spacelike-related.
    \item A lack of absolute temporal distance between spatially separated events.
\end{enumerate}
The physical picture presented by (1) is certainly strange: every body is causally isolated from every other, despite their standing in absolute spatial relations to one another.\footnote{\label{fn:Leibniz1}Speaking philosophically, this seems to be a realisation of a Leibnizian monadology. Consider e.g.:
\begin{quote}
    There is no way of explaining how a monad can be altered or changed internally by some other creature, since one cannot transpose anything in it, nor can one conceive of any internal motion that can be excited, directed, augmented, or diminished within it, as can be done in composites, where there can be change among the parts. The monads have no windows through which something can enter or leave. Accidents cannot be detached, nor can they go about outside of substances, as the sensible species of the Scholastics once did. Thus, neither a substance nor an accident can enter a monad from without. \parencite[\emph{Monadology} 7]{Leibniz}
\end{quote}
We're grateful to Oliver Pooley for suggesting this connection.} {\color{teal} (Of course, this relies on its being the case that the causality relation is tethered to the spacetime structure of Carroll gravity; one might resist this in light of some of the points which we'll go on to make below.)}
Lévy-Leblond's point here is echoed by e.g.\ \textcite[p.\ 3]{HansenPaper}: ``Particles with non-zero energy cannot move in space anymore, and for these particles there can be no interactions between spatially separated events.'' 
On the other hand, however, it seems to be resisted by e.g.\ \textcite[p.\ 9]{BergshoeffCarroll}, who argue that while it is true that a single free particle in a Carroll spacetime cannot move, this is not true for multi-particle systems in Carroll gravity, where ``only the center of mass cannot move, but [...] the separate particles can have non-trivial dynamics.''\footnote{\label{fn:Leibniz2}To continue the theme of the previous footnote: Leibniz seems to have anticipated this in the above-quoted passage, when he wrote that for composites, ``there can be change among the parts''.} {\color{teal} (\textcite{BergshoeffCarroll} also show that after gauging the Carroll algebra one can couple even a single Carrollian particle to the background gauge fields, but we won't consider this further here.)}
In particular, they claim that
\begin{quote}
    to lowest order [...] the velocity of the centre of mass is conserved, i.e.,
    \[ M_1 \frac{ \text{d} \vec{x}_1}{\text{d} t} + M_2 \frac{\text{d} \vec{x}_2}{\text{d} t} = \text{constant}.\] This implies non-trivial dynamics for the separate particles!
\end{quote}
One might be confused here: how can it be the case that in a Carroll spacetime, (a) spacelike-separated bodies are causally isolated, and yet (b) their dynamics are nevertheless coupled, according to e.g.\ the above equation? {\color{teal} (It's perhaps worth flagging that `dynamics' needn't necessarily imply evolution in time---see e.g.\ \textcite{Curiel2016-CURKDA, March2024, LinnemannRead} for conceptual discussion.)}

The underlying point here is the following. If one assumes that stress-energy propagation is timelike at the outset, then it is indeed the case that the matter dynamics in Carroll spacetimes become trivial; however, this is of course consistent with non-trivial dynamics for matter if one allows stress-energy to propagate along spacelike (as well as timelike) curves. This is nicely illustrated by our discussion of Carroll energy conditions from \S\ref{sec:Carrollspacetime}. Suppose that one adopts the strengthened dominant Carroll energy condition. Then there can be no propagation of stress-energy (i.e.,\ given this energy condition, the energy density $\rho$) off the integral curves of $\xi^a$ (i.e.,\ no spacelike propagation of stress-energy content); this, in turn, seems to underwrite the claims made by \textcite{Lévy1965} and \textcite{HansenPaper}. On the other hand, if one invokes only e.g.\ the weak Carroll energy condition, then there \emph{can} be spacelike propagation of stress-energy content, which seems better reconcilable with the conclusions of \textcite{BergshoeffCarroll}. {\color{teal} (It is on this approach that one might not wish to tether the causal relation to the timelike-separated relation in Carroll gravity.)} Indeed, if \textcite{BergshoeffCarroll} wish to maintain both (a) something akin to the strengthened dominant Carroll energy condition, alongside (b) their claims about non-trivial interactions between multi-particle Carrollian systems, then given the lack of spacelike stress-energy propagation, they will (it seems to us) need to invoke some more exotic physics.\footnote{Or even metaphysics---e.g.,\ an appeal to some version of Leibniz' doctrine of pre-established harmony (to continue the themes of footnotes \ref{fn:Leibniz1} and \ref{fn:Leibniz2}).}

Moving on, let's return to point (2) above. This is certainly also puzzling, albeit perhaps somewhat less so by virtue of its being already familiar from relativistic physics. Indeed, the lack of an absolute temporal standard is not regarded as a problem in the relativistic context so long as one can build and operationalise suitable clocks in relativistic spacetimes (for recent discussion see e.g.\ \textcite{Fletcher2013-FLELCA, Menon2018-MENCAC-3}). \emph{In principle}, one can say the same thing of Carroll spacetimes, although here again point (1) is relevant, for the `causal disconnectedness' (to return to the way of putting things from \textcite{Lévy1965}) of all bodies might stand in the way of any such operationalisation (evidently, our discussion of Carroll energy conditions will be relevant here also).



\section*{Acknowledgements}

We thank Quentin Vigneron for detailed feedback on a previous draft, and Jeremy Butterfield, Adam Caulton, Oliver Pooley, the audience of the Oxford Philosophy of Physics research seminar, and four anonymous referees, for helpful discussions and feedback. EM acknowledges financial support from Balliol College, Oxford, and the Faculty of Philosophy, University of Oxford.

\printbibliography

\end{document}